\pdfoutput=1
\documentclass[10pt,conference]{IEEEtran}

\usepackage{graphicx}
\usepackage{booktabs}
\usepackage{tabularx}
\usepackage{amsmath}
\usepackage{amssymb}
\usepackage{cite}
\usepackage{url}

\title{From Code Review Comments to Reuse in Production: An Industrial Study of Extracted Rules}

\author{
\IEEEauthorblockN{
  Mutsuha Asada\IEEEauthorrefmark{1}\IEEEauthorrefmark{2},
  Yuhei Nakasaka\IEEEauthorrefmark{2},
  Keito Tabuchi\IEEEauthorrefmark{2},
  Shinpei Hayashi\IEEEauthorrefmark{3}
}
\IEEEauthorblockA{
  \IEEEauthorrefmark{1}University of Tsukuba, Japan
}
\IEEEauthorblockA{
  \IEEEauthorrefmark{2}GA technologies Co., Ltd., Japan\\
  Email: \{m\_asada, y\_nakasaka, k\_tabuchi\}@ga-tech.co.jp
}
\IEEEauthorblockA{
  \IEEEauthorrefmark{3}Institute of Science Tokyo, Japan\\
  Email: hayashi@comp.isct.ac.jp
}
}

\begin{document}

\maketitle

\begin{abstract}
Much organization-specific code review knowledge remains embedded in comments on individual changes, and recent systems reuse it by turning those comments into rules a language model applies to later changes.
Teams must then decide which rules to keep, and the evidence at hand is how often the system's later comments name each rule.
Tracing the rules of one such system, Kleene, through 65 days of production use, we found that such a count does not show what a rule contributed.
A rule attribution indicates neither whether the source content was preserved nor whether the comment requested a change.
Of 56 clauses in 12 rules that later comments named as their reason, 2 were explicit in a source review thread, 38 added conditions to a corresponding passage, and 16 had no passage we could locate; 45 of 79 further comments naming those rules requested no change, 27 of them while stating that the rule applied.
On 20 constructed code changes, a reviewer model given only the unedited source thread met the detection criteria on 8 of the 10 that require a report, against 10 of 10 with the extracted rule, a difference whose confidence interval includes zero.
Stage counts locate where rules drop out but not why: of 464 rules deployed while the rule-based reviewer subagent ran in their repository, 259 (55.8\%) were named, 195 (42.0\%) produced a comment that was not withheld, and 94 (20.3\%) one with a recorded code change; the largest reduction occurred before any rule was named.
Nor did the deployment-time properties we tested settle the choice in advance: retaining the top-ranked half preserved 53.8\% of the rules that later produced a non-withheld comment, against 49.2\% at random, a difference whose confidence interval includes zero.
Judging extracted rules therefore calls for recording each stage separately and comparing each rule against the threads it was extracted from.
\end{abstract}


\section{Introduction}
\label{sec:introduction}

\subsection{Industrial Motivation}
\label{sec:intro-motivation}

Developers expect code review to transfer knowledge and keep changes consistent with the norms of their team and project, not only to find defects~\cite{bacchelli2013expectations,sadowski2018modern,davila2021mcr,tufano2025slr}.
Much of what a reviewer checks is therefore specific to an organization, and that knowledge exists in a form that resists reuse: a comment on one change, in one thread, by one reviewer.

Because a general-purpose language model does not reliably reflect those conventions, recent systems persist them as rules derived from past review comments~\cite{bitsai2025,qodo2026ruleminer,cursor2026bugbotlearning}.

Building such a pipeline raises a selection problem that those reports do not settle: not every past comment is worth preserving, and each retained rule adds to the cost of maintenance and retrieval.
The design below is how the system we study answers it, not a contribution we evaluate.
Our pipeline therefore applies a \emph{reproduction gate} (Section~\ref{sec:approach}): it is designed to keep a review thread only when a review model, given the same change but not the human review comment, fails to produce an equivalent comment.

Passing the reproduction gate does not establish what the rule will do after deployment: the situation it describes may rarely recur, or may recur without the reviewer subagent retrieving the rule.

A pipeline therefore has evidence for extracting a rule but not for keeping it, and a record of how often a rule was named does not supply that evidence either.
One rule, extracted from a request to extract repeated logic into a shared definition, was later named by a comment that stated the rule applied and then requested nothing, because a \texttt{TODO} and a linked issue already documented the duplication (Section~\ref{sec:rq3}).
A count of how often each rule was named does not separate such a comment from one that demanded a change.

\subsection{Research Questions}
\label{sec:intro-rqs}

We therefore ask what happens to extracted review knowledge after it is deployed, and what evidence a team needs in order to decide which rules to keep:

\begin{itemize}
  \item \textbf{RQ1: What happens in production to rules produced by the extraction pipeline?}
  \item \textbf{RQ2: Do properties of the extracted rules predict their downstream use?}
  \item \textbf{RQ3: What content is preserved or altered when review comments are extracted into rules and reused?}
\end{itemize}

RQ1 and RQ2 work from the counts a deployment records, and RQ3 reads the comments behind those counts to interpret what they record.
To answer them we analyze the production records of Kleene, an LLM-based review system we built and operate at GA technologies.
The author of a change runs Kleene on their own work before opening a pull request (Section~\ref{sec:approach}).
We chose this setting because operating the system ourselves lets us observe each stage separately, including the candidates the reviewer subagent withdrew, which a study of the comments developers were shown cannot see.

\subsection{Contributions}
\label{sec:intro-contributions}

This paper makes three contributions.

\begin{itemize}
  \item An analysis of rule attributions showing that they distinguish neither comments that request a change from those that request none nor content from the source thread from content added during extraction, with the unedited source thread as a baseline for what extraction adds.
  \item A staged trace of 464 deployed rules locating where they drop out (259 named, 195 presented, 94 with a recorded code change) and which stage telemetry cannot explain, telling a team what to record before retiring a rule.
  \item A within-repository test of deployment-time rule selection, showing that pooled metrics can mistake a repository's exposure for rule quality and that such properties give only limited ranking.
\end{itemize}

Section~\ref{sec:approach} describes Kleene, Section~\ref{sec:empirical-study} the empirical study and its threats, and Section~\ref{sec:discussion} the lessons.

\section{Related Work}
\label{sec:related-work}

\subsection{Mining Rules from Development History}
\label{sec:mining-rules}

Prior work derives reusable rules from development history, mining recurring changes to learn bug fixes~\cite{nguyen2010fixwizard}, systematic edits~\cite{meng2013lase}, project-specific bug patterns~\cite{higo2020ammonia}, and fix patterns suggested at industrial scale~\cite{bader2019getafix}.
Amazon CodeGuru promotes only a minority of mined clusters into static-analysis rules after expert curation~\cite{codeguru2023rulemining}, and industrial deployments report that the value of automated findings depends on how they reach developers~\cite{sadowski2015tricorder,sadowski2018static}.
BitsAI-CR starts from rules rather than mined code changes: a curated taxonomy of 219 rules drives a fine-tuned checker whose output a second model validates, and an Outdated Rate, approximating from subsequent changes how often developers act on a rule's comments, informs the retirement of rules that are precise but ignored~\cite{bitsai2025}.

Scoring individual rules by what developers do with their comments is therefore not new.

\subsection{Extracting Review Rules in Practice}
\label{sec:rw-practice}

Turning human review feedback into persistent guidance is now a recurring industrial practice: teams extract review knowledge from pull-request history and reuse it in later AI reviews~\cite{estie2026reviewer,m3tech2025reviewer,enechange2026reviewinsights,stationwise2026reviewer}, and commercial tools expose similar mechanisms~\cite{qodo2026ruleminer,codeant2026repolearnings,cursor2026bugbotlearning}.

They differ in the signal that decides which feedback becomes persistent guidance: how often an observation recurs~\cite{estie2026reviewer}, whether it generalizes~\cite{m3tech2025reviewer}, whether developers acted on it~\cite{qodo2026ruleminer}, an LLM or human judgment of value~\cite{stationwise2026reviewer,codeant2026repolearnings}, and reactions to the system's own comments~\cite{cursor2026bugbotlearning}.

Kleene uses a different signal within a broader pipeline: after screening review threads for generalizability, value, and duplication (Section~\ref{sec:rule-extraction}), it re-runs a review model offline on the historical change with the human comment hidden and is designed to keep the thread only if the comment is not reproduced.
That signal comes from an offline rerun on the historical change, not from an observation that a human reviewer commented on something the system had not flagged.

\subsection{Industrial Evaluations of LLM-Based Code Review}
\label{sec:industrial-llm-code-review}

Learned and retrieval-based approaches automated code refinement, comment generation, and recommendation from past reviews before LLM-based systems, and later work examined their failures, noisy data, and retrieval augmentation~\cite{tufano2021towards,tufano2022pretrained,li2022codereviewer,hong2022commentfinder,tufano2024strengths,liu2025noisy,hong2025rag}.

What makes a human review comment useful has itself been studied~\cite{bosu2015useful}; recent studies of LLM-based review in industry differ mainly in the outcome they treat as evidence of usefulness: production use of generated edits~\cite{frommgen2024resolving}, resolution and closure time~\cite{cihan2025automated}, whether reviewers forwarded comments to authors~\cite{olewicki2026impact}, whether code changes followed~\cite{tantithamthavorn2026rovodev}, and how developers responded~\cite{goldman2025resolved,lin2026agentic}.
Closest to our setting, AutoCommenter flags violations of Google's written best-practice documents and estimates a comment-resolution rate near 40\%~\cite{vijayvergiya2024autocommenter}; it applies existing documents rather than rules extracted from review threads.

These studies assess LLM-based code review by what developers did with a generated comment, or by the code changes that followed: a comment that names a rule and is followed by a code change shows a recorded response, not whether what carried forward is what the human reviewer wrote.

We therefore trace the source review thread, the extracted rule, the later review comment, and the recorded outcome together, and compare each extracted rule with the source review threads from which it was extracted.

\section{LLM-Based Self-Review with Kleene}
\label{sec:approach}

Kleene is an LLM-based code review system that we designed and have operated at GA technologies.

\subsection{Review Process with Self-Review}
\label{sec:target-system}

\begin{figure}[t]
  \centering
  \includegraphics[width=\columnwidth]{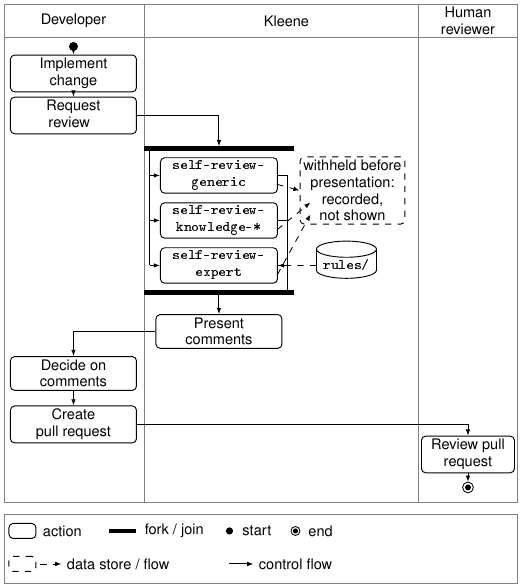}
  \caption{Self-review in the review process, interactive mode.}
  \label{fig:kleene-flow}
\end{figure}

Kleene reports potential issues in a developer's changes before they reach human review.
We call this \textbf{self-review} because the developer who wrote the change, rather than a separate reviewer, requests it and reads the result.
Whereas many LLM-based review tools run inside the pull-request workflow~\cite{github-copilot-review,coderabbit-review,qodo-review}, Kleene runs on the author's own machine, implemented as an Agent Skill~\cite{agent-skills} inside the developer's coding agent.

In Figure~\ref{fig:kleene-flow}, the developer requests a review, at which point Kleene forks three types of reviewer subagents, each started without the orchestrator's conversation history.
\texttt{self-review-generic} works from the source code alone;
\texttt{self-review-knowledge-*} draws on external sources through Model Context Protocol servers~\cite{mcp};
and \texttt{self-review-expert} reads the deployed rule files, shown as the \textsf{rules/} data store: it receives the triggers of every rule in the repository, selects the ones it judges relevant, and opens only those bodies.
Each subagent produces \emph{candidate review comments}, writing each to a file as soon as it finds it so a long run does not lose its early candidates.
The orchestrator rewrites each candidate into the shorter \emph{presented review comment} passed to the presentation stage, merging duplicates into one such comment while each keeps its own record, so a presented comment restates its candidates rather than repeating them.
Not every candidate reaches the developer: the dashed edge carries candidates a reviewer subagent produced and then withdrew, which Kleene records but never shows.
A reviewer subagent marks a candidate \texttt{suppressed} when it cannot answer yes to all three questions it must ask before reporting: whether the issue could become a production fault or future debt, whether it affects others writing the same pattern, and whether the grounds would persuade the developer.
Withholding is therefore its own decision, not a side effect of merging duplicates.
In an interactive session that runs to completion, the developer records a decision on each presented review comment before human review; those decisions, with what auto mode records on its own, are what we analyze.

In auto mode Kleene applies the repair itself and records the outcome without any decision by the developer.
Decisions are written when the developer finishes reviewing, so a session that ends before then leaves its presented comments in the \texttt{pending} state, and a \texttt{pending} comment may later expire.

\subsection{Rule Extraction Pipeline}
\label{sec:rule-extraction}

Rule extraction derives reusable rules from the review comments left on merged pull requests, where a \emph{review thread} is one comment thread on a diff: an initial \emph{review comment} at a file and line, with any replies.
A large language model (LLM) carries out each stage below except deployment.

\begin{figure}[t]
  \centering
  \includegraphics[width=0.72\columnwidth]{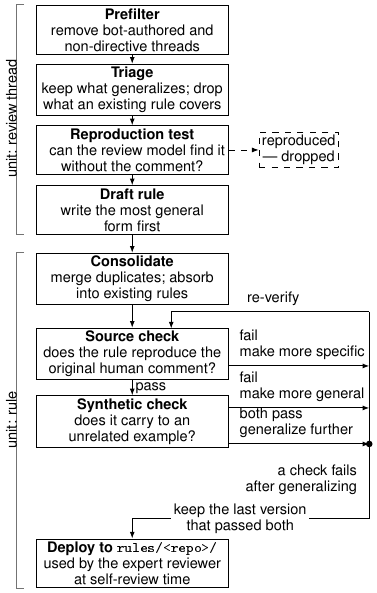}
  \caption{What each extraction stage decides.}
  \label{fig:rule-extraction-pipeline}
\end{figure}

Figure~\ref{fig:rule-extraction-pipeline} states what each stage decides, and Table~\ref{tab:rule-populations} reports how many review threads and rules reached each stage; early logging was incomplete, and Section~\ref{sec:study-method} explains why those counts are not nested.
Every stage except deployment relies on an LLM judgment against written criteria, not a deterministic classifier, and \textsf{Prefilter} and \textsf{Triage} run before the reproduction test, which is therefore neither the first nor the only stage that discards threads.

\textsf{Reproduction test} asks whether a \emph{reproduction-test model} reproduces the human review comment without an added rule.
The repository is checked out at the commit the comment was written against, the comment is hidden, and the model reviews the diff with a neutral prompt and no access to connectors or history, in up to three attempts; an LLM judge then decides whether its comment and the human one describe the same issue~\cite{zheng2023judge}.

If the reproduction-test model reproduces the human review comment in none of its attempts, the thread is used to draft a \emph{candidate rule}.
Candidate rules making the same point are consolidated, one is dropped when a deployed rule covers it, and each remaining one is checked twice by \texttt{self-review-expert} with the human comment hidden: against the diff it was written on, and against a synthetic example generated from the rule text alone.
These checks drive an iterative rewrite: a draft failing the first is made more specific and one failing the second more general, so the loop keeps the most general wording that passes both.
Neither check uses a negative case on which the reviewer should remain silent despite receiving the rule, so the checks establish that \texttt{self-review-expert} comments on an example other than the source when given the rule, not that it stays silent when it should.
Only the candidate rules leaving this loop are written into a repository-specific directory under \texttt{rules/} and become \emph{deployed rules}; one review thread may yield several, and one rule may draw on several threads.

\begin{table}[t]
  \caption{One chain, with excerpts translated from Japanese.}
  \label{tab:chain-example}
  \centering
  \small
  \begin{tabularx}{\columnwidth}{@{}lX@{}}
    \toprule
    \textbf{Source review thread} & ``it seems good to specify a maximum and a minimum for the number of \texttt{ids[]},'' with a snippet setting \texttt{maxItems} and \texttt{minItems} on an OpenAPI schema \\
    \addlinespace[2pt]
    \textbf{Deployed rule} & keeps that instruction and adds a reason: without an upper bound ``queries, loops and external calls that grow with the element count turn directly into resource consumption, response latency and denial of service'' \\
    \addlinespace[2pt]
    \textbf{Later comment} & two months later, in another branch, on a diff adding an error notification inside a per-record loop: ``the notification added here creates exactly this per-element external call'' \\
    \bottomrule
  \end{tabularx}
\end{table}

Table~\ref{tab:chain-example} shows one \emph{chain}: a source review thread, the rule extracted from it, and a later comment that named that rule and was recorded as \texttt{fixed}.

Whenever \texttt{self-review-expert} produces a candidate review comment it names the deployed rule it treats as the basis, and we link that record to whether the comment was presented and to the decision recorded on it.

\section{Empirical Study}
\label{sec:empirical-study}

\subsection{Study Setting and Data}
\label{sec:study-method}

We studied Kleene in the Engineering organization of GA technologies.
The source review comments span February 28, 2024 to August 14, 2026, the extraction pipeline ran between July 1 and August 19, 2026, and the production analysis uses sessions executed between July 7 and September 9, 2026, inclusive.
The 3,579 threads come only from merged pull requests in the 19 repositories where developers chose to run the extraction, not across the whole organization.
These periods overlap: rules were extracted and deployed while Kleene was already in production, so there is no training period followed by a disjoint evaluation period.
Rules were rarely retired: of 544 rules ever deployed, 37 had been deleted by the end of the window, and 36 of those before it began.

During this 65-day window, 942 self-review sessions were run under 46 distinct GitHub account identifiers, across 555 pull requests in 42 repositories; 18 of those repositories held deployed rules and 16 held at least one rule that had an opportunity to run.
Kleene was experimental and its use encouraged but not required, so these counts reflect its adopters, not the organization's size.
Each developer decides whether to run self-review and in which mode, interactive or auto (668 and 273 sessions; one has no mode recorded).
Our data combine the rule-extraction records, Kleene's production metrics, and the GitHub pull requests identifying each rule's source review.
Models are used below for production comment generation (not pinned, so we do not compare models), reproduction-test generation (three variants), RQ2 annotation (\texttt{gpt-5.6-terra}), RQ3 generation, scoring and clause coding (\texttt{claude-opus-5}), and drafting its checklist items (model not recorded).

We call a candidate review comment \textbf{rule-attributed} when it names a deployed rule as its basis, and that recorded relation an \emph{attribution}; below we write \emph{candidate} for a candidate review comment and \emph{presented comment} for one the operational definition below classifies as presented.
Three further notions must be kept apart from attribution.
A \emph{review opportunity} is a session in which \texttt{self-review-expert} ran while the rule was deployed in that repository (Section~\ref{sec:method-rq1}); the number of opportunities is not \emph{applicability}, which is whether the change satisfies the conditions under which the rule is relevant, and which we never observe.
We also do not observe \emph{retrieval}, whether the reviewer subagent opened the rule's body: neither the triggers it selected nor the bodies it read are written to telemetry.
An attribution is what the records show, and it does not establish that the rule was necessary for the comment that names it.

\begin{table}[t]
  \caption{Populations entering the analysis.}
  \label{tab:rule-populations}
  \centering
  \small
  \begin{tabular}{llr}
    \toprule
    \textbf{Stage} & \textbf{Unit} & \textbf{$n$} \\
    \midrule
    Human review threads collected & thread & 3,579 \\
    Passed prefilter & thread & 2,230 \\
    Passed triage & thread & 961 \\
    Blind reproduction test recorded & thread & 974 \\
    Not reproduced in any attempt & thread & 627 \\
    Draft rule recorded & thread & 567 \\
    \midrule
    Rules ever deployed & rule & 544 \\
    Overlapped production window & rule & 508 \\
    \bottomrule
  \end{tabular}
\end{table}

Records from the first weeks of the extraction period are incomplete and the early pipeline did not always enforce the stage order, so the stage counts in Table~\ref{tab:rule-populations} are not strictly nested: some recorded reproduction tests have no linked triage record or follow a triage rejection, some triaged threads have no recorded test, and a few drafts come from threads not marked \texttt{model\_missed}.
The reproduction test records one result per review thread, and the result we use is \texttt{model\_missed}: in none of the attempts did the reproduction-test model produce a comment that the LLM judge matched to the human review comment, so the thread proceeds to drafting.
These tests were not a controlled experiment: developers ran the extraction workflow as part of their ordinary work, so we read a \texttt{model\_missed} result as the output of the procedure as it ran at the time, not a determination under one fixed model.

Kleene records no separate event for showing a comment to the developer, so we \emph{operationally define} a candidate as \textbf{presented} when it was neither marked \texttt{suppressed} by the reviewer subagent nor assigned the legacy \texttt{dismissed} state; either state makes it a \emph{withheld candidate}.
Under this definition, 7,697 of the 13,287 candidates generated in the window were presented.
Throughout the paper, then, \emph{presented} is shorthand for a candidate the records do not mark as withheld: no telemetry confirms that the developer saw it, and the definition admits candidates left \texttt{pending} or \texttt{expired}.
In the comment-level counts we use only candidates whose named rule matches one deployed in the same repository: of the 1,842 candidates \texttt{self-review-expert} produced in the window, 43 named no resolvable rule, leaving 1,799 attributed to 258 rules.
The rule-level counts come from a per-rule table that records an attribution wherever the attributing session ran, giving 259 rules; the extra rule was named only from a session in another repository, and that comment was \texttt{suppressed}, so the difference changes no later count.

\subsection{RQ1: Downstream Use of Deployed Rules}
\label{sec:rq1}

\subsubsection{Motivation}

Counting how many rules were deployed is easy, but a single adoption rate merges the separate events traced below and so would not show where in that sequence the number of rules falls most.

\subsubsection{Method}
\label{sec:method-rq1}

The cohort for RQ1 is every rule deployed for at least part of the production window, identified from the repository history rather than from the extraction ledger.
Of the 464 rules in that cohort with at least one review opportunity, we traced 410 back to a source review thread, meaning the ledger holds an entry recording both a \texttt{model\_missed} result and the rule file's deployment.
For the other 54 we found no such entry under the rule's name, so we report the RQ1 rates over every deployed rule and restrict claims about a rule's origin to the chains below.
A rule's observation interval runs from the later of its first deployment and the window's start to the earlier of its deletion and the window's end; events count only inside it.

Because each rule is deployed to one repository, a \emph{review opportunity} for a rule is a session in that repository, within its observation interval, in which \texttt{self-review-expert} was launched, whether or not it produced a candidate.
For each rule we trace four successive points: the rule is deployed, a candidate review comment names it, that candidate is classified as presented, and the presented comment carries a recorded code-change outcome, counting each rule at most once at each point; attributions are resolved as described in Section~\ref{sec:study-method}.
Because \texttt{fixed} may record a developer's code change or an automated repair in auto mode, we read it as a recorded code-change outcome rather than as developer acceptance.
The orchestrator merges duplicates only in the text the developer sees, so each candidate keeps its own record and decision and one code change can be recorded against several: of the 190 attributed comments with a recorded code change counted below, 58 share a file and line with another reviewer's comment in the same session that also carries one.
Self-review also runs before the pull request, so in interactive mode \texttt{fixed} records an author acting on a comment about their own unpushed work, not a separate reviewer's comment followed by a code change; our rates are not comparable with resolution rates reported for pull-request review systems.
Our primary rates use rules with at least one review opportunity as the denominator, because a rule is named from a session in its own repository, with the one exception noted in Section~\ref{sec:study-method}.
As a descriptive check we summarize interactive presented comments separately, because a \texttt{fixed} outcome in auto mode may record a repair Kleene made itself; there a comment carries a recorded decision when its state is \texttt{fixed}, \texttt{declined} or \texttt{not\_selected}.

\subsubsection{Results}
\label{sec:results-rq1}

Of the 508 rules in the production cohort, 464 had at least one review opportunity.
Among these, 259 (55.8\%) were named by at least one candidate, 195 (42.0\%) produced at least one presented comment, and 94 (20.3\%) at least one presented comment with a recorded \texttt{fixed} outcome.
Counting only rules with at least one comment carrying a recorded decision lowers the presented figure to 155 rules (33.4\%).
The largest single reduction therefore occurred before any attribution: 205 rules (44.2\% of the 464) were never named by any candidate review comment.

The 205 rules that no candidate named did not lack review opportunities: across all 464 rules the median was 27 (IQR 16--44), and named rules had more than unnamed ones (median 34 against 27).
Whether any change a given rule applies to ever appeared in those sessions is not recorded, and the subagent's rule selection is not logged, so applicability and retrieval cannot be distinguished in our telemetry.
One cause can be ruled out: all but 43 of the 1,842 candidates \texttt{self-review-expert} produced in the window named a resolvable rule.
Even under a null model ignoring deployment dates and opportunity counts, many rules would receive none: for a repository holding $N$ rules over which its $C$ attributions fall independently and uniformly, the expected number receiving zero is $N(1-1/N)^{C}$; summed over the 16 repositories that is 130 of the 464 rules (28.0\%) against the 205 (44.2\%) observed.
The remaining gap is consistent with the concentration: the 50 most-named rules take 62.5\% of all attributions.

Over the window, \texttt{self-review-expert} produced 1,799 candidates whose named rule matched a deployed rule, of which 885 were presented and 190 received a \texttt{fixed} outcome.
The interactive share is 139 of those 190 comments and 77 of the 94 rules.
Counting only the 132 such comments whose code change no other reviewer's comment shares, 73 of the 94 rules still had at least one such comment.

The remaining 914 candidates were withheld before presentation, 906 of them \texttt{suppressed} by the reviewer itself and the other 8 carrying the legacy \texttt{dismissed} state.
Withholding is thus the most frequent outcome for a rule-attributed candidate in this deployment: suppression by the reviewer subagent alone accounts for 50.4\% of the 1,799.
Withheld candidates keep the text the reviewer subagent wrote and span 208 of the 258 rules named in the window;
for 63 of those 208 rules every rule-attributed candidate was withheld, which is why only 195 rules have a presented comment.
These counts describe what the reviewer subagent produced and withheld, not whether it was right to withhold.

Linking each recovered source review thread to the later comments that named the rule extracted from it and received a \texttt{fixed} outcome yields 180 chains, spanning 86 rules, 88 source review threads, and 177 later comments.
A chain records only that a rule extracted from the review of one change was named by a comment recorded \texttt{fixed} on another; it does not establish that the original reviewer's point was carried through.

In 158 of the 180 chains the account that ran the later session differs from the source pull request's author, and in 122 it differs from that author and from every person who commented on the source review thread.
In 91 of those 122 chains the later session ran in interactive mode, where a developer rather than Kleene made the code change.
No chain links a rule to its own source pull request.
Eight of the 94 rules with a recorded code-change outcome, and 13 of the 190 comments carrying one, fall outside these chains: no source review thread was linked to them under the rule's name.

The chains rest on linking records, not on checking that the later comment makes the same point as the source thread, so we examined one in detail.
In it, a human reviewer argued that a job error should keep the same retry budget as a transient error because a later deploy could still resolve it, and the deployed rule asks whether a failing condition could be resolved by another path within the retry window.
Fifty-one days later, a candidate named that rule on code that logged an unmapped external value and wrote a null result, accepted downstream and never retried; the recorded code change raised a typed error.
Not every comment naming that rule carried a code change: 7 of its 47 comments did.

\textbf{Answer to RQ1.}
Of the 464 rules with at least one review opportunity, just over half were named by a candidate, two in five produced a presented comment, and one in five a presented comment with a recorded code change, so the largest reduction occurs before attribution.

\subsection{RQ2: Rule Properties and Downstream Use}
\label{sec:rq2}

\subsubsection{Motivation}

If only some deployed rules go on to produce comments classified as presented, a team maintaining a rule set would like to identify those rules in advance.
The pipeline records several properties of every rule at the moment it is created, and a natural policy is to keep the rules those properties rank highest.
RQ2 asks whether that policy would have worked.

\subsubsection{Method}
\label{sec:method-rq2}

We use the same 464 rules from 16 repositories, with three binary outcomes matching the points RQ1 traces: whether a rule was ever named, ever produced a presented comment, and ever produced one with a recorded code change.

We take each rule's properties from the body as it stood at that rule's first deployment commit, recovered for all 464 rules including the eleven edited after deployment.
We defined four structural properties and five semantic dimensions before the analysis.
The structural properties are the number of triggers, the logarithm of the body's length, whether the body contains a fenced code block, and whether it contains a concrete example.
One semantic dimension is nominal: the rule's \emph{concern}, which is correctness, testing, maintainability, design, or operational.
The other four are ordinal, each placing the rule on a three-point scale: \emph{how far a reviewer must look} to decide whether the rule applies (the changed code alone, elsewhere in the same repository, or knowledge outside it); \emph{how general the rule's main statement is} (a specific technical pattern, a policy for a kind of structure, or a design principle); \emph{how specifically it prescribes what to do} about a violation (an explicit check, a judgment left to the reviewer, or a principle only); and \emph{how narrowly it is bound to a technology} (one named technology, a family of them, or none).
Codex CLI running \texttt{gpt-5.6-terra} annotated each rule in a fresh process from the body and rubric alone, never seeing outcomes.

Because semantic labels are useful only if applied consistently, one author independently annotated an 80-item repository-stratified sample drawn with a fixed seed, blind to the model's label and to outcomes.
We prespecified Cohen's $\kappa \geq 0.40$ between the author's and the model's labels for a semantic dimension to enter the primary models; one below that threshold enters only a joint sensitivity analysis.

We evaluate with leave-one-repository-out cross-validation, holding out each of the 16 repositories in turn and fitting all preprocessing on the rest, so the question is whether outcomes observed elsewhere can rank rules in a repository with no outcome history.
We fit L2-regularized logistic regression with $\lambda=1.0$ fixed and no hyperparameter or feature search.
Four models use only what is known at deployment: a prevalence-only baseline, a \emph{structural model} over the four structural properties, a \emph{semantic model} over the dimensions that met the reliability threshold, and their combination.
Three further models add two measures of a rule's \emph{production exposure}, the logarithm of its review-opportunity count and the length of its observation interval: exposure alone, structural plus exposure, and structural plus semantic plus exposure.
Exposure is observed only after deployment, so it cannot select rules in advance; we include it to see whether downstream use tracks a rule's review-opportunity count, which varies little within a repository and so limits the power of a within-repository test.

Our primary metric is \emph{within-repository pairwise concordance}: within each repository we form all positive--negative rule pairs and measure the pair-weighted proportion for which the model gives the positive rule a higher out-of-fold prediction, ties counting 0.5, so 0.5 is chance.
We use it rather than pooled ROC-AUC because folds differ in intercept and prevalence, so pooling would mix within-repository ranking with differences between repositories.
A repository contributes only when it holds at least one rule with the outcome and one without, leaving 13, 15, and 13 of 16 for the attributed, presented, and recorded-outcome labels.
Confidence intervals come from a bootstrap that resamples repositories 2,000 times with the out-of-fold predictions held fixed, so no model is refit, and we treat discrimination as detectable only when the interval excludes 0.5.
As a post-hoc interpretation we also retain the top 50\% by predicted score within each repository and compare the fraction of positives retained with the expectation under random retention of the same number, 49.2\% rather than 50\% because a repository with $n$ rules keeps $\max(1,\lfloor n/2 \rfloor)$ of them, and ties at the cut are broken by input order.

\subsubsection{Results}
\label{sec:results-rq2}

On the 80-item validation set only the rule's concern met our prespecified threshold of $\kappa \geq 0.40$, with $\kappa=0.674$ (95\% CI [0.546, 0.787]); the four ordinal dimensions fell below it, with unweighted $\kappa$ from 0.166 to 0.349 and quadratic-weighted $\kappa$ from 0.230 to 0.367, and are excluded from the models.
The concern is therefore the only semantic predictor in the primary analysis.

The structural model provided detectable but limited within-repository discrimination only for whether a rule later produced a presented comment.
Its concordance was 0.561 (95\% CI [0.522, 0.628]), compared with 0.500 for the prevalence-only baseline.
For the attributed label the estimate was 0.556 (95\% CI [0.470, 0.593]) and for the recorded-outcome label 0.566 (95\% CI [0.416, 0.656]), both intervals including 0.500, so we found no comparable evidence of discrimination there.

Neither the concern nor production exposure improved on the structural model for the presented label: adding the concern lowered concordance to 0.482 and adding exposure lowered it to 0.536, and the bootstrap interval for each difference excludes zero.
The concern is distributed very differently across repositories, so a fit from the others need not transfer.
Exposure on its own showed no detectable discrimination for any outcome, and a sensitivity analysis with all five semantic dimensions gave a similar concordance (0.559 versus 0.561).
Because four dimensions missed our reliability threshold, this does not show that semantic properties are generally uninformative.

Applying the out-of-fold structural-model scores as a policy keeping the top 50\% within each repository retained 105 of the 195 rules that later produced a presented comment (53.8\%), against 49.2\% expected for randomly keeping the same number.
The full-sample difference was 4.6 percentage points, but its repository-cluster bootstrap interval included zero (95\% CI [$-0.3$, $+9.5$] percentage points), and the policy still discarded 90 such rules.

A model given only production exposure appeared more predictive under pooled metrics, reaching a pooled ROC-AUC of 0.701 for the recorded-outcome label despite a within-repository concordance of 0.497.
Review-opportunity counts explain this gap: they vary almost entirely between repositories and very little within one (all 33 rules in one repository had 77 review opportunities, and all 29 in another had five), so exposure largely encoded a rule's repository.

\textbf{Answer to RQ2.}
Under the top-half pruning policy we evaluated we found no clear evidence of improved retention over random within-repository selection, although the structural properties did give detectable within-repository discrimination for presentation, and exposure's stronger pooled association reflected differences between repository deployments rather than among rules within one.

\subsection{RQ3: What Rule Reuse Preserves and Changes}
\label{sec:rq3}

\subsubsection{Motivation}

RQ1 and RQ2 leave a team to judge a rule from evidence recorded after deployment, and the count of how often a rule was named is the most readily available such evidence.
Both treat every attribution as the same kind of event, which the example in Section~\ref{sec:introduction} shows it is not.
RQ3 therefore asks what a rule-attributed comment actually says, and what the extracted rule adds beyond the source review threads from which it was extracted.

\subsubsection{Method}
\label{sec:method-rq3}

\smallskip\noindent\textit{Tracing individual attributions.}
Among the rules named by the 1,799 resolvable candidates, the population is the set of deployed rules whose record contains both a presented comment and a suppressed candidate, a condition on recorded state alone.
145 rules meet that condition.
We took 10 of those 145 rules in a fixed hash order and added 2 we had already read closely, giving 12 rules; for each we read one presented comment that requested a change and one suppressed candidate, and all 12 candidates requested none.
Because we selected these rules for having both kinds of comment, they support statements about what these rules contain, not rates over all deployed rules.

For these 12 rules we read all 103 comments that named them: the 24 above and 79 others.
Extraction may consolidate several review threads into one rule, so for each rule we recovered every source review thread the ledger records: eight have one and four have two.
Segmenting each rule body into clauses gave the 56 clauses those comments named as a stated reason.
For every clause that a later comment named as its stated reason, we coded the clause \emph{explicit} when a source comment states it, \emph{added} when a source comment states part of it and the clause carries further conditions or steps, and \emph{unlocated} when no passage in any source review thread corresponds to it.
A model performed both the clause segmentation and the coding; Section~\ref{sec:threats-construct} reports an author's later rule-level check.

We performed two additional checks.
First, we asked whether \emph{explicit}, \emph{added} and \emph{unlocated} covered the relationships to the source seen in the other 79 comments.
For every comment we read, we recorded for each clause it named whether that clause gave the comment's reason or was only quoted, and whether the comment gave reasons not stated in the rule.
Second, without seeing the segmentation, one author stated for each rule why its suppressed candidate requested nothing and how the rule relates to its source review threads.

\smallskip\noindent\textit{Comparing supplied knowledge.}
To ask whether the same review behavior could be obtained from the source review thread, we held the review task fixed and varied only the knowledge supplied to a \emph{reviewer model}: none, the unedited source review thread, or the deployed rule.

\textbf{Rule selection and case construction.}
We used only rules whose body states an exception, because one of the two case pairs below requires a change on which it holds.
We matched rule bodies against a pattern for exception wording, set aside those used while developing the design, and drew eight of the rest in a fixed hash order.
One author read the matched text of those eight and confirmed five to state an exception; we kept those five and dropped the other three without replacement.
Each came from a single source review thread, so that condition has one thread to supply.

For each of the five rules we constructed four code changes from real files at real commits, arranged as two matched pairs, giving 20 cases.
One pair varies whether the rule's trigger holds; the other whether the exception written in the rule holds.
In each pair the first change is a positive case the criteria require the review to report, and its partner alters only the premise that pair varies, so the review should report nothing.
Within a pair everything else is identical, and the written case record states what was altered and what was held fixed.

\textbf{Conditions and execution.}
Each of the 20 cases was reviewed under each of the three conditions, giving 60 runs.
Each run is a fresh \texttt{claude-opus-5} process holding only its own prompt, with the same tool set and no access to any other run's output or to the scoring criteria.
Everything outside the knowledge block and the few words framing it---the diff, the context files, the instructions and the output schema---is byte-identical across the three conditions.
The source-thread condition fills that block with every comment body verbatim, replacing reviewer names with pseudonyms, and changes the framing only as far as needed to refer to review comments rather than rules.
The no-knowledge condition keeps that framing and leaves the block empty, so the reviewer model is told that rules are in use and then given none, rather than working without such framing.

\textbf{Scoring.}
Before running the experiment, we wrote \emph{scoring criteria} for each case: the statements the review must make, and the changes it must not demand.
One author checked those criteria against the case, then labeled every run through a form showing only the output, the diff and that case's criteria, with the condition name and machine scores withheld.
The form presents the criteria as one yes-or-no checklist item at a time; a model drafted those items from the criteria.
Blinding is incomplete, because an output sometimes states that it was given no rules and we did not alter outputs.

For each case, the criteria specify whether detection requires all of the listed statements or any one of them.
A run counts as a \emph{detection} when the listed statements are satisfied under that all-or-any requirement, and a statement counts as satisfied only when every element of it appears in the output.
In one positive case, for instance, the required statement has three elements: that the added property duplicates a server-owned value set, that it does so by comparing a status field against a fixed integer, and which server-side state that integer stands for.
An output describing the duplication but not identifying that state was not a detection.
A run counts as \emph{demanding a prohibited change} when it requests a change the case's criteria list as one that must not be demanded.
Scoring uses the output's content alone: we do not use the rule an output names or its attribution field, so a run given no rule is scored by the same standard as one given the deployed rule.

We designed this comparison after seeing the RQ1 and RQ2 results, so it is exploratory; we report the first of three repetitions, the one an author labeled.

\subsubsection{Results}
\label{sec:results-rq3}

\smallskip\noindent\textit{The 12 rules we traced contain content we could not find in their source review threads.}
Across the 12 rules, 56 distinct clauses were named as a stated reason by at least one later comment: 2 of those clauses were explicit, 38 added content beyond a corresponding passage, and 16 were unlocated.
All 12 rules had a named clause that went beyond its source, counting added and unlocated clauses, and 9 contained at least one unlocated clause; restricting both counts to the 10 rules drawn in hash order gives 10 of 10 and 7 of 10.
The added content includes procedural material: it tells the reviewer model to evaluate side effects before acting, to allow for a case the source thread never mentions, or to treat a condition as out of scope.
Extraction is expected to add material: the verification loop of Section~\ref{sec:rule-extraction} makes a draft more specific when it fails on its source and otherwise generalizes it until a check fails. What this trace adds is that later comments name that material.
In a separate assessment made without seeing the segmentation, one author judged 11 of the 12 rules to correspond in part to their source review threads, with content added during extraction, and found no corresponding source passage for the remaining rule.
These are counts of what the rule text contains; they do not say whether the added material improved the rule.

Table~\ref{tab:chain-example} gives an example: we found no passage in that rule's source review thread corresponding to the clause quoted there, and that clause is the one the later comment named.
The thread asks for two schema keywords, whereas the later comment reasons about cost per element, the content extraction added.

The case in Section~\ref{sec:introduction} shows the two decisions coming apart inside one comment.
The comment named the clause asking that sites with the same trigger, arguments and side effects be pulled into one definition, recording that the rule applied.
It then requested no change, because the state the rule guards against---that a reader cannot tell whether the two sites merely look alike or express the same concept---``is resolved by the \texttt{TODO} and the issue link.''
No clause of that rule says an acknowledged duplicate may remain, so this attribution records a judgment that the rule applied while requesting nothing.

The suppressed candidates show the same split from the other side.
Of the 12 suppressed candidates, the author traced the decision to request nothing to a condition or exception written in the rule in 6 cases, to reasons outside it in 5, and to both in 1.
When we instead classified why no change was requested, 5 of the 12 gave deferral as the reason, the rule having applied and the comment having said so.

The other 79 comments show the same split: 32 requested a change and 2 made a request conditional on something the comment could not check.
Of the 45 that requested nothing, 17 gave non-applicability as the reason and 27 said the rule did apply: 14 deferred the change and 13 reported its objective already satisfied.
The one remaining comment quoted the rule without giving either reason.
Across the 103, a named clause served as the comment's reason in 91 and was only quoted in 12, and 98 also gave reasons not stated in the rule.

\smallskip\noindent\textit{The reviewer model given only the source review thread met most of the scoring criteria.}
In both knowledge-supplied conditions the reviewer model receives the source thread or rule body directly, rather than the trigger list it selects from in production, so this compares knowledge content, not retrieval.
Table~\ref{tab:rq3-conditions} summarizes all 60 runs.
Each detection count covers the 10 positive cases, and each prohibited-change count the 5 cases of one pair type: the 5 whose trigger condition no longer matches, and the 5 on which the rule's own exception holds.
On the two cases detected only with the extracted rule, the source-thread output described the underlying problem but left out one required element: which server-side state a fixed integer stood for in one case, and the records whose preconditions were rewritten in the other.
The outputs there differed in how completely they stated the problem.
A prohibited-change count includes every request the criteria forbid, whatever reason the output gives, so it is not a count of technically wrong requests.

\begin{table}[t]
  \caption{What each knowledge condition produced under the case-specific scoring criteria.}
  \label{tab:rq3-conditions}
  \centering
  \small
  \begin{tabular}{@{}lccc@{}}
    \toprule
     & Complete & \multicolumn{2}{c}{Change the criteria forbid} \\
    \cmidrule(l){3-4}
     & detection & trigger does & exception \\
    Knowledge & (10 cases) & not match & holds \\
    \midrule
    None                 & 3/10  & 3/5 & 1/5 \\
    Source thread & 8/10  & 2/5 & 4/5 \\
    Extracted rule       & 10/10 & 2/5 & 2/5 \\
    \bottomrule
  \end{tabular}
\end{table}

Detection outcomes were nested: every case detected under no knowledge was also detected with the source thread and with the extracted rule, and every case detected with the source thread was also detected with the extracted rule.
The source-thread condition satisfied the detection criteria in five cases, spanning three rules, that the no-knowledge condition missed, with no case in the opposite direction.
The clearest instance is a one-sentence source review thread asking a colleague to write \texttt{opportunity} out rather than abbreviate it.
Given that thread unedited, the reviewer model produced the required comment on a block parameter named \texttt{bk} and a local variable named \texttt{phone\_num}, which we wrote into two constructed cases and which the thread does not mention; with no knowledge supplied, neither was reported.
The extracted-rule condition satisfied the detection criteria in two cases, spanning two rules, that the source-thread condition missed, again with no case in the opposite direction.
A rule-clustered bootstrap over the five rules gives $+50.0$~percentage points for the source-thread condition over the no-knowledge condition, 95\%~CI $[+10.0, +90.0]$, and $+20.0$~points for the extracted rule over the source thread, 95\%~CI $[0.0, 40.0]$, including zero.

Prohibited change requests were not nested, and every pairwise interval includes zero.
The two pair types nevertheless behave differently: on the five cases where the rule's exception holds, the source-thread condition demanded the prohibited change on 4 and the extracted-rule condition, whose body states that exception, on 2.

The four prohibited change requests made under the extracted-rule condition are not all failures to recognize that the rule did not apply.
In two the output said the rule did not apply, or that its exception held, yet recommended a similar change on other grounds; in a third the same demand appeared under all three conditions; in the fourth it rested on a different problem.
These cases show that judging whether a rule applies is distinct from deciding whether to recommend a change.

\textbf{Answer to RQ3.}
The 12 rules we traced contain procedural content beyond what their source review threads state, and later comments name it as their reason, including comments that request no change.
On the 20 constructed cases the reviewer model given only the unedited source thread produced the required comment on code that thread does not mention, and the extracted-rule condition met the criteria in two further cases, a difference whose interval includes zero.
Some comments stated that the rule applied and still requested nothing, and some constructed-case outputs requested a change after stating it did not apply, so neither attribution counts nor change-request counts establish what the rule contributed.

\subsection{Threats to Validity}
\label{sec:threats}

\subsubsection{Construct Validity}
\label{sec:threats-construct}

Each downstream stage is an operational definition, not a direct observation: an attribution is recorded because the reviewer subagent names a rule as its basis, so a rule no comment named may have been ignored, consulted and found inapplicable, or consulted silently, and \texttt{fixed} records a code change, not acceptance or correctness.

Three further constructs rest on judgments rather than telemetry.
In RQ3 a model segmented the rule bodies and coded each clause; after the blind judgments of Section~\ref{sec:method-rq3}, one author compared each rule's clause table with its source review threads and judged the coding supported for 11 of the 12 rules, recording verdicts per rule, so no clause-level agreement can be reported.
A \texttt{model\_missed} result is the recorded output of the reproduction procedure, whose records keep the judge's verdict and rationale but not the comments it compared.
In RQ3, eight of the twenty cases take their exceptions or remedies from the rule text, and one author confirmed six from the codebase alone.
On the other two the scoring measures conformance to the rule's own boundary rather than to an independently established fact; both are exception cases, so neither is one of the two on which the extracted-rule condition detected and the source-thread condition did not.
Building the cases and their scoring criteria with the rule in view may bias the comparison toward the extracted-rule condition.

\subsubsection{Internal Validity}
\label{sec:threats-internal}

Kleene evolved during the observation period---reviewer composition, knowledge sources and the pipeline all changed---and neither the production model nor the one under test was assigned independently of time or repository, so we do not read temporal or model differences as causal.

We construct no matched set of rules from alternative pipelines, so the study does not isolate the reproduction gate's contribution, and RQ3 cannot say which part of a rule produced the differences.

\subsubsection{Conclusion Validity}
\label{sec:threats-conclusion}

Each cell holds at most ten cases drawn from five rules, so the intervals are wide, we report paired case counts alongside them, and an interval including zero is not evidence of equivalence.
One author produced all 405 human statement labels over the 60 runs, and 83 were entered through a control marking a run's remaining statements absent at once; 82 are change statements, so it barely affects the detection counts.
\texttt{claude-opus-5} also scored all three repetitions, making those scores partly self-assessment; they put the no-knowledge condition lowest in every repetition, while the extracted rule and the source thread differ by at most one case.
The 10 positive cases give 30 runs across the three conditions, carrying 42 detection checks the author and the model both scored.
They agreed on 31; in 10 of the 11 disagreements the model was more permissive, so we report the author-labeled repetition rather than pooling.

In the tracing analysis, correspondence was judged against the threads we could recover, so absence means we did not locate one, not that the reviewer never held the belief.

\subsubsection{External Validity}
\label{sec:threats-external}

This is a single deployment in one Japanese real-estate technology company, and the identifiers we count are GitHub accounts, not verified individuals.
The telemetry records employees' use of an internal tool and the quoted excerpts are our translations; both uses were approved within the company, and the reported results carry no personal identifiers.
Because Kleene is author-side and selects from triggers before reading rule bodies, the rates may differ for systems that inject all rules into context, retrieve by embedding, or present comments to a separate reviewer; the RQ3 cases were constructed, not sampled.

\section{Lessons Learned}
\label{sec:discussion}

No maintenance policy was evaluated in production, so the lessons below are implications, not tested procedures.

\textbf{Compare each extracted rule against the source review threads it was built from.}
Asking only whether a pipeline's rules were named cannot show what the rewriting contributed: with no baseline every attribution is credited to the pipeline.
In RQ3 the source-thread condition met the detection criteria on 8 of 10 cases against 10 of 10, and demanded the prohibited change on 4 of the 5 exception cases against 2 (Table~\ref{tab:rq3-conditions}); both intervals include zero.
Our comparison supplies the relevant source thread directly; in production it would also have to be retrieved from the review history, whereas Kleene selects among deployed rules by their triggers.
A source-thread baseline is therefore available to any organization holding that history, and the rewriting may still be worth its cost for reasons we did not measure.

\textbf{Record why a rule was never named before concluding that it is unnecessary.}
For the 205 rules (44.2\%) that no candidate named, our telemetry cannot tell whether that reflects non-recurrence, non-applicability, retrieval failure or correct silence, because Kleene records only the rule a comment names, not which bodies were opened; and a null model distributing attributions uniformly within each repository predicts 28.0\% unnamed.
Record which rules the reviewer subagent consulted and judged inapplicable: that separates \emph{retrieval} from its applicability judgment and from \emph{attribution}, though actual applicability remains unobserved.

\textbf{Treat extracting a rule and keeping a rule as two separate decisions.}
The chains in RQ1 show that some rules the extraction pipeline produced were later named by review comments carrying a recorded code change, but RQ2 found that among deployed rules the properties recorded at deployment ranked only weakly.
Deploy extracted rules provisionally and reassess them using post-deployment evidence beyond attribution and code-change counts; running a change with the rule and with its source thread is one way to obtain it, untested here as a retention policy.

\textbf{Evaluate learned rules within their deployment context.}
Raw reuse rates can make differences between deployments look like differences between rules: RQ2's exposure-only model looked predictive when repositories were pooled yet ranked at chance within them, because a rule in a repository with many sessions has more chances at each outcome.
Compare within a deployment, or account for exposure, before reading reuse rates as evidence of rule quality.

\section{Conclusion}
\label{sec:conclusion}

We traced rules extracted from historical code review comments through 65 days of production use.
Reading the comments attributed to 12 rules exposed a limit of attribution counts: a comment naming a rule does not reveal whether its stated reason came from the source review material or from content added during extraction, nor whether it requested a change.
The stage counts locate where rules drop out but not why: two in five of the 464 rules produced a presented comment, and one in five one with a recorded code change; in 122 of the 180 chains the later session ran under an account that took no part in the source review thread.
Deployment-time properties did not clearly improve on random retention.
Judging this kind of reuse therefore calls for instrumenting each stage separately and comparing each rule against the threads it was extracted from.

\section*{Data Availability}

We cannot release the data behind this study.
The review comments it starts from carry confidential information about products GA technologies operates, and that information runs through everything derived from them: the rule bodies, the production records, and the files the RQ3 cases were built on.
The excerpts quoted here can be shown; the corpus cannot.

\section*{Acknowledgments}

We thank the Engineering organization at GA technologies for the usage capacity behind the rule-extraction workflow and for their feedback.
Generative AI assisted with wording and with checking this text against our records; the design, analyses, and conclusions are ours, and we verified every reference.

\bibliographystyle{IEEEtran}
\bibliography{references}

\end{document}